\documentclass[11pt,a4paper]{article}
\pdfoutput=1
\usepackage{jheppubnohead}
\usepackage[utf8]{inputenc}
\usepackage{amsmath}
\usepackage{epsfig}
\usepackage{graphicx}
\usepackage{amssymb}
\usepackage{tabularx}
\usepackage[normalem]{ulem} 
\usepackage{booktabs} 
\usepackage{subfig}

\providecommand{\proarrow}[0]{\rightarrow}

\providecommand{\dif}[0]{\mathrm{d}}

\providecommand{\proname}[2]{#1 \proarrow #2}

\providecommand{\abs}[1]{\left\lvert #1 \right\rvert}

\providecommand{\abss}[1]{\left\lvert #1 \right\rvert^2}

\providecommand{\mire}[1]{{\rm Re} \left[ #1 \right]}
\providecommand{\miim}[1]{{\rm Im} \left[ #1 \right]}

\providecommand{\order}[1]{{\cal O} \left( #1 \right)}
\providecommand{\torder}[1]{{\cal O} \bigl( #1 \bigr)}

\providecommand{\cpm}[0]{\mathcal{M}}
\providecommand{\pep}[0]{{\bf G}^{(1,{\rm p})}}
\providecommand{\pmed}[0]{p_0}
\providecommand{\mmed}[0]{M}
\providecommand{\emed}[0]{E_0}
\providecommand{\treal}[0]{\theta_{\rm R}}
\providecommand{\tim}[0]{\theta_{\rm I}}

\newcommand{\be}{\begin{equation}}
\newcommand{\ee}{\end{equation}}
\newcommand{\bea}{\begin{eqnarray}}
\newcommand{\eea}{\end{eqnarray}}

\title{CP violation in mixing and oscillations for leptogenesis with quasi-degenerate neutrinos}
\author[]{J.~Racker}

\affiliation[]{Instituto de Astronom\'{\i}a Te\'orica y Experimental (IATE), Universidad Nacional
de C\'ordoba (UNC)~- Consejo Nacional de Investigaciones Cient\'{\i}ficas y T\'ecnicas
(CONICET), Laprida 854, X5000BGR,  C\'ordoba, Argentina.\\
Observatorio Astron\'omico de C\'ordoba (OAC), Universidad Nacional de C\'ordoba (UNC), Laprida 854, X5000BGR, C\'ordoba, Argentina.
}

\emailAdd{jracker@unc.edu.ar}

\abstract{We study the sources of CP violation for baryogenesis models with quasi-degenerate neutrinos. Our approach is to use the renormalized propagator in a quantum field theory model of neutrino oscillations, paying close attention to unitarity requirements. From the probabilities of lepton number violating processes obtained in this way, we derive a source term for the time evolution of the lepton asymmetry. The source term has contributions that can be identified with CP violation from mixing, oscillations and interference between both. Given that this source term does not involve processes with unstable particles in the initial or final states, neither does it require to calculate number densities of neutrinos, no subtraction of real intermediate states must be performed. In equilibrium the source term is null, as demanded by unitarity and CPT invariance, due to a cancellation between the terms coming from CP violation in mixing and oscillations. The calculations are done in a simple scalar toy model, and the resummed propagator is diagonalized at first order in the decay widths over the mass difference. We also comment on the effect of the interference term, which is mild at the order we work, but seems to become more important with increasing degeneracy. 
}
\begin{document}

\maketitle
\section{Introduction}

Among models of baryogenesis involving new physics around or below the TeV scale, those with two or more exotic particles nearly degenerate in mass are an interesting option. Notably, in the type I seesaw model it is possible to have low scale leptogenesis, either in the freeze-out of Majorana neutrinos with $\order{1~{\rm TeV}}$ masses, i.e. via the so-called resonant leptogenesis mechanism~\cite{pilaftsis03}, or in the freeze-in of much lighter neutrinos, i.e. baryogenesis via neutrino oscillations, also called ARS leptogenesis~\cite{akhmedov98, asaka05}. Resonant leptogenesis has been studied carefully with different formalisms (see~\cite{Dev:2017wwc} for a comprehensive review). More recently ARS leptogenesis has received a lot of attention, in part because it has been discovered that it can be probed and even tested in some regions of parameter space in planned experiments (we refer the reader to the review~\cite{Drewes:2017zyw} for explanations and references). Some research efforts in the last few years have been in understanding the more complex region of parameter space involving neutrinos with masses in the intermediate mass range of several tens to hundreds of GeV, which requires a proper understanding of the role of the helicity~\cite{Eijima:2017anv, Ghiglieri:2017gjz, Hambye:2017elz, Ghiglieri:2017csp, Eijima:2018qke, Abada:2018oly, Ghiglieri:2018wbs, Klaric:2020lov}.  

In all these models involving quasi-degenerate particles, the CP even phase required to have CP violating processes comes from the absorptive part of loop amplitudes or from oscillating phases due to the coherent propagation of different mass eigenstates. The interplay of these sources of CP violation has been analyzed in detail under different formalisms and approximations in~\cite{Dev:2014laa, Dev:2015wpa, Dev:2014wsa, Kartavtsev15} (see also~\cite{Fidler:2011yq, garbrecht11, garny11, Garbrecht:2014aga}).  Following a semi-classical approach, a fully flavour-covariant set of transport equations involving a matrix  of  number  densities was derived in~\cite{Dev:2014laa, Dev:2015wpa}. In that formalism it is necessary to subtract the real intermediate state contributions to some scattering processes to avoid violation of unitarity, and this issue seems even more subtle than with classical Boltzmann equations (BE), as it is necessary to account for thermal corrections when considering off-diagonal flavour correlations.  
Instead, the analysis in~\cite{Dev:2014wsa} and~\cite{Kartavtsev15} is based on the Kadanoff-Baym formalism of non-equilibrium thermal field theory. Although several of the conclusions in these works are compatible, including that the oscillation and mixing sources can contribute additively to the final asymmetry, an interference term between mixing and oscillations was found in~\cite{Kartavtsev15} while not in~\cite{Dev:2014laa, Dev:2015wpa, Dev:2014wsa} (to understand this issue it might be important to note that these works make different approximations according to the washout regime they focus on). 

It seems clear that the treatment of CP violation for leptogenesis models with quasi-degenerate neutrinos is not a trivial subject and has actually been discussed over some decades now. Therefore different looks at this problem can be useful. In this regard, the effective Hamiltonian formalism is a simple approach that has been applied successfully to the  phenomenology of CP violation in meson decays~\cite{Zyla:2020zbs}. This method was used in~\cite{Liu:1993ds, covi96II} to calculate the CP asymmetry in the decay of heavy particles with arbitrary mass splittings that mix, which elucidated various issues. Still, some points remained open for a complete implementation of the method to baryogenesis. Namely, how do the time dependent or time integrated asymmetries enter in transport equations?, and related to this point, which initial state(s) should be considered in the CP asymmetries calculated in~\cite{Liu:1993ds, covi96II}? It is also key to determine the relevant conditions imposed by unitarity, which might not be trivial given the non-Hermiticity of the effective Hamiltonian. The purpose of this work is to address these questions. We will follow a quantum field theory approach that, up to some point and under certain approximations, can be matched to the effective Hamiltonian formalism. 

The work is organized as follows: In section~\ref{sec:toymodel} we calculate the renormalized propagator in a scalar toy model, diagonalize it under certain approximations and verify the relevant unitarity conditions. The corresponding amplitudes for the lepton number violating processes are used in a quantum field theory model of oscillations in section~\ref{sec:osc}, where we also verify that unitarity is satisfied for the probabilities obtained in this way. From those probabilities we derive, in section~\ref{sec:st}, a source term which only involves stable particles as asymptotic states and discuss the different contributions. Finally, in section~\ref{sec:con} we summarize the main results and comment on possible directions for future work.

\section{Renormalized propagator and unitarity}
\label{sec:toymodel}
The issues we want to study in this work can be captured in a simple scalar toy model commonly used for this type of purposes in several of the references given above (in particular in~\cite{Kartavtsev15}). It consists of one complex and two real scalar fields, denoted by $b$ and $\psi_i$ ($i=1,2$), respectively. In a basis where the mass matrix of the real scalars is diagonal,
the Lagrangian is given by
\begin{equation}
\label{eq:lag}
\mathcal{L} = \frac{1}{2} \partial^\mu \psi_i \, \partial_\mu \psi_i - \frac{1}{2} \psi_i M_i^2 \psi_i +  \partial^\mu \bar b \, \partial_\mu b - m^2 \, \bar b b - \frac{h_i}{2} \psi_i \, b b - \frac{h_i^*}{2} \psi_i \, \bar b \bar b - \frac{\lambda}{2\cdot 2}(\bar b b)^2 \; .
\end{equation}

The $b$-particles will subsequently be called ``leptons'', since they play in this toy model the analogous role that leptons play in standard leptogenesis, and for simplicity their mass $m$ will be neglected. The lepton charge is broken by the cubic Yukawa interaction terms involving the $\psi_i$, to be called ``neutrinos'' in what follows. The last term is a quartic interaction which does not change lepton number but might be used as a way to localize the leptons and satisfy the conditions to have oscillations~\cite{Beuthe01}, but we will not make explicit use of it. 

The one-loop renormalized inverse propagator matrix ${\bf G}^{-1}$ is given by 
 \begin{equation}
 i {\bf G}^{-1}(p^2) = p^2 {\bf 1} - {\bf M^2}(p^2),
 \end{equation}
 with
 \begin{equation}
 {\bf M^2}(p^2)= 
\begin{pmatrix}
M_1^2 + \Sigma_{11}(p^2)  & \Sigma_{12}(p^2) \\
\Sigma_{21}(p^2)  & M_2^2 + \Sigma_{22}(p^2) 
\end{pmatrix},
\end{equation}
and
\begin{eqnarray}
\Sigma_{ii}(p^2)&=&\frac{\abss{h_i}}{(4\pi)^2} \left[1 + \ln \frac{p^2}{M_i^2} - \frac{p^2}{M_i^2} - i \pi \theta(p^2)\right], \\
\Sigma_{12}(p^2)&=&\Sigma_{21}(p^2)=\frac{\mire{h_1^* h_2}}{(4\pi)^2} \left[\frac{M_2^2 \ln \frac{p^2}{M_1^2} - M_1^2 \ln \frac{p^2}{M_2^2} - p^2 \ln \frac{M_2^2}{M_1^2}}{M_2^2-M_1^2} - i \pi \theta(p^2)\right] .
\end{eqnarray}
 Here $\theta$ is the step function, which will be omitted in the following calculations given that we will always evaluate these expressions for $p^2 > 0$ (the symbol $\theta$ will be used in the rest of the paper to denote the quantity defined below). We have used the following renormalization conditions:
 \begin{eqnarray}
 \mire{\Sigma_{ii}(M_i^2)} &=& 0, \quad {\rm for} \; i=1,2\, , \\
 \mire{\Sigma_{12}(M_i^2)} &=& 0, \quad {\rm for} \; i=1,2\, , \\
 \frac{\dif \Sigma_{ii}}{\dif p^2}\Bigr\rvert_{p^2=M_i^2} &=& 0, \quad {\rm for} \; i=1,2\, .
 \end{eqnarray}

In order to use a quantum field theory model for oscillations in the next section, we proceed to diagonalize the propagator matrix. To simplify the analysis we will perform the diagonalization at first order in the quantities
\begin{equation}
\label{eq:eta}
\eta_{ij} \equiv \frac{\abs{h_i h_j}/(4 \pi)^2}{M_2^2 - M_1^2} .
\end{equation} 
Doing so, we will not be able to study the highly degenerate case $M_2 - M_1 \sim \Gamma_{1,2}$, which is left for future work (note that e.g. in~\cite{anisimov05} a similar expansion is performed to make a careful comparison of different procedures to obtain the CP-asymmetry in the decays of quasi-degenerate Majorana neutrinos). 
At first order in $\eta_{ij}$, ${\bf G} = {\bf G}^{(1)} + \torder{\eta_{ij}^2}$, with
\begin{equation}
{\bf G}^{(1)}(p^2) = i 
\begin{pmatrix}
1 & \theta \\
-\theta & 1
\end{pmatrix}
\begin{pmatrix}
(p^2 - M_1^2 - \Sigma_{11})^{-1} & 0 \\
0 & (p^2 - M_2^2 - \Sigma_{22})^{-1} 
\end{pmatrix}
\begin{pmatrix}
1 & -\theta \\
\theta & 1
\end{pmatrix},
\end{equation}
and
\begin{equation}
\label{eq:theta}
\theta \equiv \frac{\Sigma_{12}}{M_2^2 - M_1^2} \, .
\end{equation}

The poles $\cpm_{a,b}^2$ of the diagonal propagator matrix are given by the solutions to
\begin{eqnarray}
\cpm_a^2 - M_1^2 - \Sigma_{11} (\cpm_a^2) = 0  \quad &\Longrightarrow& \quad \cpm_a^2 = M_1^2 - i M_1 \Gamma_1 + \order{h_1^4},\\
\cpm_b^2 - M_2^2 - \Sigma_{22} (\cpm_b^2) = 0 \quad &\Longrightarrow& \quad \cpm_b^2 = M_2^2 - i M_2 \Gamma_2 + \order{h_2^4},
\end{eqnarray}
with $\Gamma_i \equiv \tfrac{\abss{h_i}}{16 \pi M_i}$ the total decay widths. Next, it will also be convenient to make an expansion around the complex poles (see e.g.~\cite{fuchs16} for more details on the pole structure of the propagator matrix). This  yields ${\bf G}^{(1)}(p^2) = \pep (p^2)+ \torder{(p^2-\cpm_{a,b}^2)^{0},h_i^4}$, with 
\begin{equation}
\pep (p^2) = i 
\begin{pmatrix}
1 & \theta \\
-\theta & 1
\end{pmatrix}
\begin{pmatrix}
(p^2 - \cpm_a^2)^{-1} & 0 \\
0 & (p^2 - \cpm_b^2)^{-1} 
\end{pmatrix}
\begin{pmatrix}
1 & -\theta \\
\theta & 1
\end{pmatrix}.
\end{equation}

Unitarity and CPT invariance place strong requirements for transition probabilities $\abss{A(\proname{i}{j})}$, which are crucial for a correct implementation of baryogenesis mechanisms~\cite{sakharov67, Weinberg:1979bt, Kolb:1979qa} (an improper implementation may led to generation of spurious asymmetries). For the scalar model of our study and given that we will not consider the unstable neutrinos as initial or final states, the key relation imposed by unitarity and CPT invariance (at zeroth order in the $\lambda$ coupling and sixth order in the Yukawa couplings) is
\begin{equation}
\abss{A(\proname{b b}{b b})} + \abss{A(\proname{b b}{\bar b \bar b})} = \abss{A(\proname{b b}{b b})} + \abss{A(\proname{\bar b \bar b}{b b})},
\end{equation}
and therefore
\begin{equation}
\label{eq:unit}
\Delta \abss{A(\proname{\bar b \bar b}{b b})} \equiv \abss{A(\proname{\bar b \bar b}{b b})}  -  \abss{A(\proname{b b}{\bar b \bar b})} = 0 . 
\end{equation}
For a propagator matrix ${\bf G}$ the invariant matrix elements are given by
\begin{eqnarray}
\mathcal{M}(\proname{\bar b \bar b}{b b}) & = i & \sum_{j,k} h_j^* \, {\bf G}_{j k} \, h_k^* \, ,\\
\mathcal{M}(\proname{b b}{\bar b \bar b}) & = i & \sum_{j,k} \, h_j {\bf G}_{j k} \, h_k \, .   
\end{eqnarray}
It is straightforward to demonstrate that the unitarity requirement~\ref{eq:unit} is satisfied plugging in these expressions the exact one-loop resummed propagator $\bf{G}$ or, at the corresponding order, any of the approximations ${\bf G}^{(1)}$ or $\pep$.
Note that there are also one-loop vertex contributions at the same order in the Yukawa couplings, but they cancel independently in~\ref{eq:unit} (see e.g.~\cite{roulet97}) and will not be considered in this work, since they are not enhanced by the quasi-degeneracy of the neutrinos. 

\section{Oscillations and unitarity}
\label{sec:osc}
Keeping the terms up to first order in $\eta_{ij}$ in the propagator $\pep$, the invariant matrix elements for the lepton number violating processes become 
\begin{eqnarray}
\label{eq:ML}
-\mathcal{M}(\proname{\bar b \bar b}{b b}) & = & \left( h_1^{* 2} - 2 h_1^* h_2^* \theta \right) \Delta_1 +   \left( h_2^{* 2} + 2 h_1^* h_2^* \theta \right) \Delta_2 \, , \\
-\mathcal{M}(\proname{b b}{\bar b \bar b}) & = & \left( h_1^2 - 2 h_1 h_2 \theta \right) \Delta_1 +   \left( h_2^2 + 2 h_1 h_2 \theta \right) \Delta_2 \, ,
\end{eqnarray}
with
\begin{equation*}
\Delta_j \equiv \frac{1}{p^2 - M_j^2 + i M_j \Gamma_j} \, .
\end{equation*}
These amplitudes can be used in a quantum field theory model of oscillations. We will consider an external wave packet model~\cite{1963AnPhy22, PhysRevD.48.4310} following the detailed review and analysis of~\cite{Beuthe01}. In this model the initial and final states of a given process are described by localized wave packets. Assuming that the factors related to coherence and localization which could destroy oscillations can be neglected, the probabilities of the lepton number violating processes are given by
\begin{eqnarray}
\label{eq:AL}
\abss{A}(L)  &=&  N \left|\left( h_1^{* 2} - 2 h_1^* h_2^* \theta \right)  e^{-i \left(M_1 - i \frac{\Gamma_1}{2}\right) \frac{\mmed L}{\pmed}} + \left( h_2^{* 2} + 2 h_1^* h_2^* \theta \right)    e^{-i \left(M_2 - i \frac{\Gamma_2}{2}\right) \frac{\mmed L}{\pmed}}  \right|^2 , \notag \\
\abss{\bar A}(L)  &=&  N \left|  \left( h_1^2 - 2 h_1 h_2 \theta \right) e^{-i \left(M_1 - i \frac{\Gamma_1}{2}\right) \frac{\mmed L}{\pmed}} +   \left( h_2^2 + 2 h_1 h_2 \theta \right) e^{-i \left(M_2 - i \frac{\Gamma_2}{2}\right) \frac{\mmed L}{\pmed}}  \right|^2 .
\end{eqnarray}
Here, to simplify the notation we have defined $A \equiv A(\proname{\bar b \bar b}{b b})$ and $\bar A \equiv A(\proname{b b}{\bar b \bar b})$,  $L$ is the distance between the production and decay of the neutrinos that mediate these processes, while $\mmed \equiv (M_1+M_2)/2$ and $\pmed$ are the average values of the mass and momentum of the neutrinos, respectively. We have integrated over solid angle and the normalization constant $N$ is going to be determined below (see~\cite{Beuthe01} for the case of stable neutrinos). These expressions for the probabilities are valid up to first order in $(M_2^2-M_1^2)/(2 \pmed^2)$ and, under the approximations we have made, can be matched to an effective Hamiltonian approach~\cite{Beuthe01}.
For the following discussion it will be more convenient to change from distance $L$ to time $t$ via the relation $\tfrac{\mmed L}{\pmed}=\tfrac{t}{\gamma}$, with $\gamma \equiv \emed/\mmed$ the Lorentz factor and $\emed$ the average energy (i.e. $\tfrac{\mmed L}{\pmed}$ is the classical proper time of propagation).

Unitarity and CPT invariance imply that, for a given initial state, $\sum_j \abss{A(\proname{i}{j})} = \sum_j \abss{A(\proname{\bar i}{\bar j})}$ (with the bar denoting CP conjugate states). The unitarity condition~\ref{eq:unit} that we verified in the previous section involved initial and final particles with well defined momentum, while in eqs.~\ref{eq:AL} the states (particularly the final states), have been taken as wave packets localized in space. Therefore, to verify that the probabilities in eq.~\ref{eq:AL} respect unitarity, we must perform a sum over all possible final states, i.e. an integral over $L$  (or equivalently over $t$, as noted above). We will come back to this crucial point below, but before we define some time dependent CP odd quantities to be used in the rest of our study.

As is well known, CP violation requires both, a relative CP-even phase (given by the factor $\miim{I_0 I^*_1}$ below) and a relative CP-odd phase (given by the factor $\miim{\lambda_0 \lambda^*_1}$ below). Specifically, consider two contributions to a certain amplitude, with the couplings factored into the parameters $\lambda_i$, so that $A(\proname{i}{j}) = \lambda_0 I_0 + \lambda_1 I_1$ and $A(\proname{\bar i}{\bar j}) = \lambda^*_0 I_0 + \lambda^*_1 I_1$. Then one gets 
\begin{equation}
\Delta \abss{A(\proname{i}{j})} \equiv  \abss{A(\proname{i}{j})} -  \abss{A(\proname{\bar i}{\bar j})} = - 4 \miim{\lambda_0 \lambda^*_1} \miim{I_0 I^*_1} \; .
\end{equation}
We apply this general expression to obtain the CP asymmetry $\Delta \abss{A} \equiv \abss{A} - \abss{\bar A}$ from eqs.~\ref{eq:AL}, noticing that there are two different types of CP even phases: one independent of $L$ (or $t$) in $\theta$, and an oscillating one in the exponentials  $e^{-i  M_j  t/ \gamma}$. Considering all the interferences and the source of the CP even relative phases, the CP asymmetry can be written as a sum of contributions from mixing $M$ (involving only $\theta$), from oscillations $O$ (involving only $e^{-i  M_j  t/ \gamma}$), and interference terms $I$ (involving both $\theta$ and $e^{-i  M_j  t/ \gamma}$):
\begin{equation}
\label{eq:deltaA}
 \frac{\abss{A}(t) - \abss{\bar A}(t)}{N} = M(t) + O(t) + I(t) \, ,
\end{equation}  
with
\begin{eqnarray}
\label{eq:AM}
M(t) &= & 8 \, \miim{h_1 h_2^*} \tim \left( \abss{h_1} e^{-\Gamma_1 t/\gamma} + \abss{h_2} e^{-\Gamma_2 t/\gamma}  \right), \\
\label{eq:AO}
O(t) & = & 8 \, \miim{h_1 h_2^*} \mire{h_1 h_2^*} \miim{e^{i (M_2 - M_1) t/\gamma}} e^{-\Gamma t/\gamma}, \\ 
I(t) & = &  8 \, \miim{h_1 h_2^*} \left( \abss{h_1} \miim{e^{i (M_2 - M_1) t/\gamma} \, \theta^*} - \abss{h_2} \miim{e^{i (M_2 - M_1) t/\gamma} \, \theta}  \right) e^{-\Gamma t/\gamma}, \notag \\
\label{eq:AI}
 & \simeq &  - 8 \, \miim{h_1 h_2^*} \tim \left( \abss{h_1} + \abss{h_2}  \right) \mire{e^{i (M_2 - M_1) t/\gamma}} e^{-\Gamma t/\gamma},
\end{eqnarray}
where we have written $\Gamma \equiv (\Gamma_1+\Gamma_2)/2$ and $\theta = \treal + i \tim$, so that
\begin{equation}
\tim = - \frac{\mire{h_1 h_2^*} \pi}{(4 \pi)^2 \left(M_2^2 - M_1^2\right)}\, .
\end{equation}
In the last line of eq.~\ref{eq:AI} we have neglected the terms proportional to $\treal$, which are suppressed by two powers of the Yukawa couplings compared to the $O$ term in eq.~\ref{eq:AO} and, moreover, $\treal$ vanishes exactly for $p^2=M_{1,2}^2$.

As explained above, to check whether our approach respects unitarity, we must integrate the probabilities over all times. Indeed, considering the basic integrals:
\begin{eqnarray*}
\int_0^\infty \sin \left((M_2 - M_1) t/\gamma \right) e^{-\Gamma t/\gamma} \dif t & = & \frac{M_2 - M_1}{\left(M_2- M_1\right)^2 + \Gamma^2} \; \gamma, \\
\int_0^\infty \cos \left((M_2 - M_1) t/\gamma \right) e^{-\Gamma t/\gamma} \dif t& = & \frac{\Gamma}{\left(M_2- M_1\right)^2 + \Gamma^2} \; \gamma, 
\end{eqnarray*}
it is immediate to see that 
\begin{eqnarray}
\int_0^\infty M(t) \dif t & = & 8 \, \miim{h_1 h_2^*} \tim \left( \frac{\abss{h_1}}{\Gamma_1} +  \frac{\abss{h_2}}{\Gamma_2} \right) \; \gamma, \\
\label{eq:intO}
\int_0^\infty O(t) \dif t & = & 8 \, \miim{h_1 h_2^*} \mire{h_1 h_2^*} \frac{M_2 - M_1}{\left(M_2- M_1\right)^2 + \Gamma^2} \; \gamma, \\
\label{eq:intI}
\int_0^\infty I(t) \dif t & = & \order{h^8}, 
\end{eqnarray}
where $\order{h^n}$ represents terms that are order $n$ in the Yukawa couplings $h_{1,2}$.  
  Hence 
  \begin{equation}
  \label{eq:unit3}
  \int_0^\infty  \abss{A}(t) \, \dif t = \int_0^\infty  \abss{\bar A}(t) \, \dif t + \order{h^8},
  \end{equation}
  and therefore unitarity is verified up to the order we have been working, i.e up to $\order{h^6}$. Note that the interference term, although giving an $\order{h^8}$ contribution when integrated over all times, gives an $\order{h^6}$ contribution at finite times, and in fact cancels the CP asymmetry from mixing at small times (relative to the oscillation period). Moreover, the $\order{h^8}$ terms in eq.~\ref{eq:unit3} are, to be more specific, $\order{h^4 \eta^2}$, with $\eta$ representing any of the $\order{h^2}$ quantities introduced in eq.~\ref{eq:eta} which increase with decreasing $\Delta M^2 \equiv M_2^2 - M_1^2$. It can be seen directly from eq.~\ref{eq:intO} that the oscillation term gives a contribution $\order{h^2 \eta^3}$, which could be the dominant one at $\order{h^8}$, but it is actually canceled by the interference term. Given that we have diagonalized the propagator at first order in $\eta$, we cannot make a meaningful discussion beyond lowest non-trivial order, however the cancellation just mentioned suggests that the interference term might have an important role in the highly degenerate limit, as found in~\cite{Kartavtsev15}. We will illustrate these issues related to the interference term in the next section. 
  
  A final comment may be of interest. Defining $\abss{A(\proname{\bar b \bar b}{\psi})}(t) \equiv \int_t^\infty  \abss{A}(t') \, \dif t' $ and $\abss{A(\proname{b b}{\psi})}(t) \equiv \int_t^\infty  \abss{\bar A}(t') \, \dif t' $, eq~\ref{eq:unit3} can be written as
  \begin{equation*}
  \int_0^t  \abss{A}(t') \, \dif t' + \abss{A(\proname{\bar b \bar b}{\psi})}(t) = \int_0^t  \abss{\bar A}(t') \, \dif t' + \abss{A(\proname{b b}{\psi})}(t) + \order{h^8}.
  \end{equation*}
 In this form the unitarity condition could be interpreted as involving a sum over all possible final states at a finite time $t$, with $\abss{A(\proname{\bar b \bar b}{\psi})}(t)$ and $\abss{A(\proname{b b}{\psi})}(t)$ giving the probability that the neutrinos mediating the corresponding processes have not yet decayed (one could also include lepton number conserving contributions, but they cancel due to CPT invariance).

\section{Source term}
\label{sec:st}

The time evolution of the lepton density asymmetry,  $n_L \equiv n_b - n_{\bar b}$, can be obtained from the sum of two terms, 
 \begin{equation}
 \frac{\dif n_L}{\dif t} = S(t) - W(t),
 \end{equation}
 where the source term $S(t)$ is the part which may be non-null in the absence of a lepton density asymmetry and $W(t)$ is the so-called washout term. We want to build the source term for a transport equation of the lepton asymmetry directly from the probabilities~\ref{eq:AL}, without resorting to some count of neutrino number densities. Two considerations will be important for this purpose: (i) In a classic-like approach to transport equations we can choose a small time window and consider the processes that produce and destroy leptons and antileptons. Unitarity and CPT invariance imply that the total probability of destruction processes must be the same for leptons and antileptons. Therefore the net effect of destruction processes is non-null only if there is some lepton asymmetry, i.e. destruction processes do not contribute to the source term (see e.g.~\cite{racker18} for more details and some subtle issues on this point). Then the source term can be obtained considering only the production processes. (ii) For the production of leptons and antileptons we must take into account that, e.g.~a pair of leptons produced at time $T$, might come from a process involving the annihilation of antileptons at a previous time $t$. Therefore a proper integration of the probabilities~\ref{eq:AL} over the whole history of the system must be considered.

For our purposes it is enough to consider a static universe and, to keep things simple, that all the neutrinos mediating the processes in eqs.~\ref{eq:AL} have the same average momentum $p_0$, so that momentum integrals are avoided (the procedure can be generalized to include the expansion of the universe and more realistic momentum distributions, which is work in progress). Finite density effects will also not be included. 
In order to determine the normalization constant $N$ in eqs.~\ref{eq:AL} and show how to integrate over time the probabilities, we start by looking at the density rate of lepton production at time $T$ due to antilepton annihilations, to be denoted by $\gamma(\proname{\bar b}{b})(T)$ below. At tree level, considering for the time being only the processes mediated by $\psi_1$ and using the narrow width approximation,

 \begin{equation}
 \label{eq:gbbb}
\gamma(\proname{\bar b}{b})(T) = 2 \int_0^T \gamma(\proname{\bar b \bar b}{\psi_1}) e^{-\Gamma_1 (T-t)/\gamma} \Gamma(\proname{\psi_1}{b  b})/\gamma \; \dif t \, .
 \end{equation}
 Here the first factor in the integral, $\gamma(\proname{\bar b \bar b}{\psi_1})$, is the density rate of $\psi_1$ production by antileptons, the exponential factor takes into account the fraction of $\psi_1$ produced at time $t$ that have survived at the time of interest $T$, and $\Gamma(\proname{\psi_1}{b  b})/\gamma$ is the decay rate into leptons divided by the Lorentz factor. The rate $\gamma(\proname{\bar b \bar b}{\psi_1})$ is given by $\gamma(\proname{\bar b \bar b}{\psi_1}) = n^{\rm eq}(t) \, \Gamma(\proname{\bar b \bar b}{\psi_1})/\gamma$, where $n^{\rm eq}(t)$ is the equilibrium density of a scalar particle of mass $\mmed$. Although in realistic calculations $n^{\rm eq}(t)$ would be a function of the time dependent temperature, in the examples given below for a static universe we will artificially vary $n^{\rm eq}(t)$ and equilibrium will simply correspond to constancy over time. Finally the factor 2 in eq.~\ref{eq:gbbb} comes because two leptons are produced in each process. Due to CPT invariance $\Gamma(\proname{\bar b \bar b}{\psi_1})=\Gamma(\proname{\psi_1}{b  b})$, both rates being equal to $\Gamma_1/2$ at tree level. Therefore
  \begin{equation}
\gamma(\proname{\bar b}{b})(T) = 2 \int_0^T n^{\rm eq}(t) \left(\frac{\Gamma_1}{2  \, \gamma}\right)^2 e^{-\Gamma_1 (T-t)/\gamma} \; \dif t \, .
 \end{equation}

 The normalization constant $N$ can be determined by equating the above expression to the one obtained using the corresponding tree level term of eqs.~\ref{eq:AL} (i.e. the first term of the first equation) :
 \begin{equation}
2 \int_0^T n^{\rm eq}(t) N \abs{h_1}^4 e^{-\Gamma_1 (T-t)/\gamma} \; \dif t  = 2 \int_0^T n^{\rm eq}(t) \left(\frac{\Gamma_1}{2  \, \gamma}\right)^2 e^{-\Gamma_1 (T-t)/\gamma} \; \dif t \, ,
 \end{equation}
 and therefore,  $N=1/(32 \pi \emed)^2$ (within our approximation of considering a single average energy $\emed$). Anyway, it should be noted that the unitarity relation~\ref{eq:unit3} obtained from integrating the terms in eq.~\ref{eq:deltaA} does not depend on the value of $N$.
 
Next the source term at time $T$ can be obtained from a similar time integral of the probabilities of lepton production minus antilepton production, including all the terms of eqs.~\ref{eq:AL}. In this way, using eq.~\ref{eq:deltaA}, we get
 \begin{equation}
 \label{eq:source}
 \begin{split}
 S(T) = 2 & \int_0^T \frac{n^{\rm eq}(t)}{(32 \pi \emed)^2} \, 8 \, \miim{h_1 h_2^*} \bigg\{  \theta_I \left[  \abss{h_1} e^{-\Gamma_1 (T-t)/\gamma} + \abss{h_2} e^{-\Gamma_2 (T-t)/\gamma} \right.  \bigg. \\
 & \left. - \big(\abss{h_1}+\abss{h_2}\big) \cos\big((M_2-M_1)(T-t)/\gamma\big) \, e^{-\Gamma (T-t)/\gamma} \right] \\
 & \bigg. + \mire{h_1 h_2^*}   \sin\big((M_2-M_1)(T-t)/\gamma\big) \, e^{-\Gamma (T-t)/\gamma} \bigg\} \; \dif t \; .
 \end{split}
 \end{equation}
In equilibrium, i.e. when $n^{\rm eq}(t)$ remains constant for a time period larger than the other time scales (the oscillation period and lifetimes of neutrinos), the unitarity condition~\ref{eq:unit3} ensures that $S(t)$ becomes null.
 
 For comparison, the standard calculation of the wave function contribution to the CP asymmetry in the decay of $\psi_i$, for $\eta_{ij} \ll 1$, yields
 \begin{equation}
 \epsilon_i \equiv \frac{\Gamma(\proname{\psi_i}{b b}) - \Gamma(\proname{\psi_i}{\bar b \bar b})}{\Gamma(\proname{\psi_i}{b b}) + \Gamma(\proname{\psi_i}{\bar b \bar b})} = \frac{\mire{h_i^* h_j} \miim{h_i^* h_j}}{8 \pi \abss{h_i} } \frac{1}{M_j^2 - M_i^2} ,
 \end{equation}
 where $j=2 (1)$ for $i= 1 (2)$. 
In terms of $\epsilon_i$ and $\Gamma_i$, the source term in eq.~\ref{eq:source} can be written in a form that eases comparison with the standard source term of classical BE:
\begin{equation}
 \label{eq:source2}
 \begin{split}
 S(T) = 2 & \bigg\{  \left[ \epsilon_1 \Gamma_1 n_{\psi_1}(T) + \epsilon_2 \Gamma_2 n_{\psi_2}(T)\right]/\gamma  \bigg. \\
 & - \frac{\left[ \epsilon_1 \Gamma_1^2 + \epsilon_2 \Gamma_2^2\right]}{\gamma^2}
   \int_0^T n^{\rm eq}(t) \cos\big((M_2-M_1)(T-t)/\gamma\big) \, e^{-\Gamma (T-t)/\gamma} \, \dif t \\
 & \bigg. + \int_0^T \frac{n^{\rm eq}(t)}{(32 \pi \emed)^2} 8 \, \miim{h_1 h_2^*}  \mire{h_1 h_2^*}   \sin\big((M_2-M_1)(T-t)/\gamma\big) \,e^{-\Gamma (T-t)/\gamma} \bigg\} \, \dif t \, ,
 \end{split}
 \end{equation}
 where
 \begin{equation}
 n_{\psi_i}(T) \equiv \int_0^T n^{\rm eq}(t) \frac{\Gamma_i}{\gamma} e^{-\Gamma_i (T-t)/\gamma} \, \dif t  .
 \end{equation}
Note that $n_{\psi_i}(T)$ is the solution to the differential equation
\begin{equation*}
\frac{\dif n_{\psi_i}(T)}{\dif T} = -\frac{\Gamma_i}{\gamma} \left[n_{\psi_i}(T)-n^{\rm eq}(T)\right]
\end{equation*}
with the initial condition $n_{\psi_i}(0)=0$, and therefore $n_{\psi_i}(T)$ corresponds to the number density of $\psi_i$ calculated with the classical BE in the absence of oscillations. In the standard classical treatment, appropriate for large enough mass splittings, the source term can be built considering the production of leptons by the density rates $\gamma(\proname{\psi_i}{b b})$ and $\gamma'(\proname{\bar b \bar b}{b b})$, where the prime in the second rate means that a real intermediate state subtraction must be performed to be consistent with unitarity. Subtracting the corresponding production terms for antileptons, the classical source term $S^{\rm cl}(T)$ reads  
\begin{equation}
\label{eq:sourcecl}
S^{\rm cl}(T) = 2 \, \epsilon_1 \frac{\Gamma_1}{\gamma} \left[n_{\psi_1}(T) - n^{\rm eq}(T) \right] + 2 \, \epsilon_2 \frac{\Gamma_2}{\gamma} \left[n_{\psi_2}(T) - n^{\rm eq}(T) \right] ,
\end{equation}
where the terms proportional to $n_{\psi_i}(T)$ come from the production of leptons and antileptons via $\psi_i$ decays, and the ones proportional to $n^{\rm eq}(T)$ come from the off-shell lepton and antilepton annihilations. Therefore, the contribution from mixing in the source term $S(T)$ (first line of eq.~\ref{eq:source2}), exactly matches the contribution from decays to the standard classical source $S^{\rm cl}(T)$. Both sources satisfy the unitarity requirement that their total integral over time is zero if the population of neutrinos is null at the beginning and the end.  In $S(T)$ the mixing contribution is canceled by the oscillation contribution, while in $S^{\rm cl}(T)$ the off-shell annihilations cancel the production form decays.  Indeed, next we show with some plots  that the lepton asymmetry obtained integrating $S(t)$, tends to the one obtained from $S^{\rm cl}(T)$, as the mass splitting $\Delta M/\mmed \equiv (M_2-M_1)/\mmed$ increases.

\begin{figure}[!t]
\centerline{\protect\hbox{
\epsfig{file=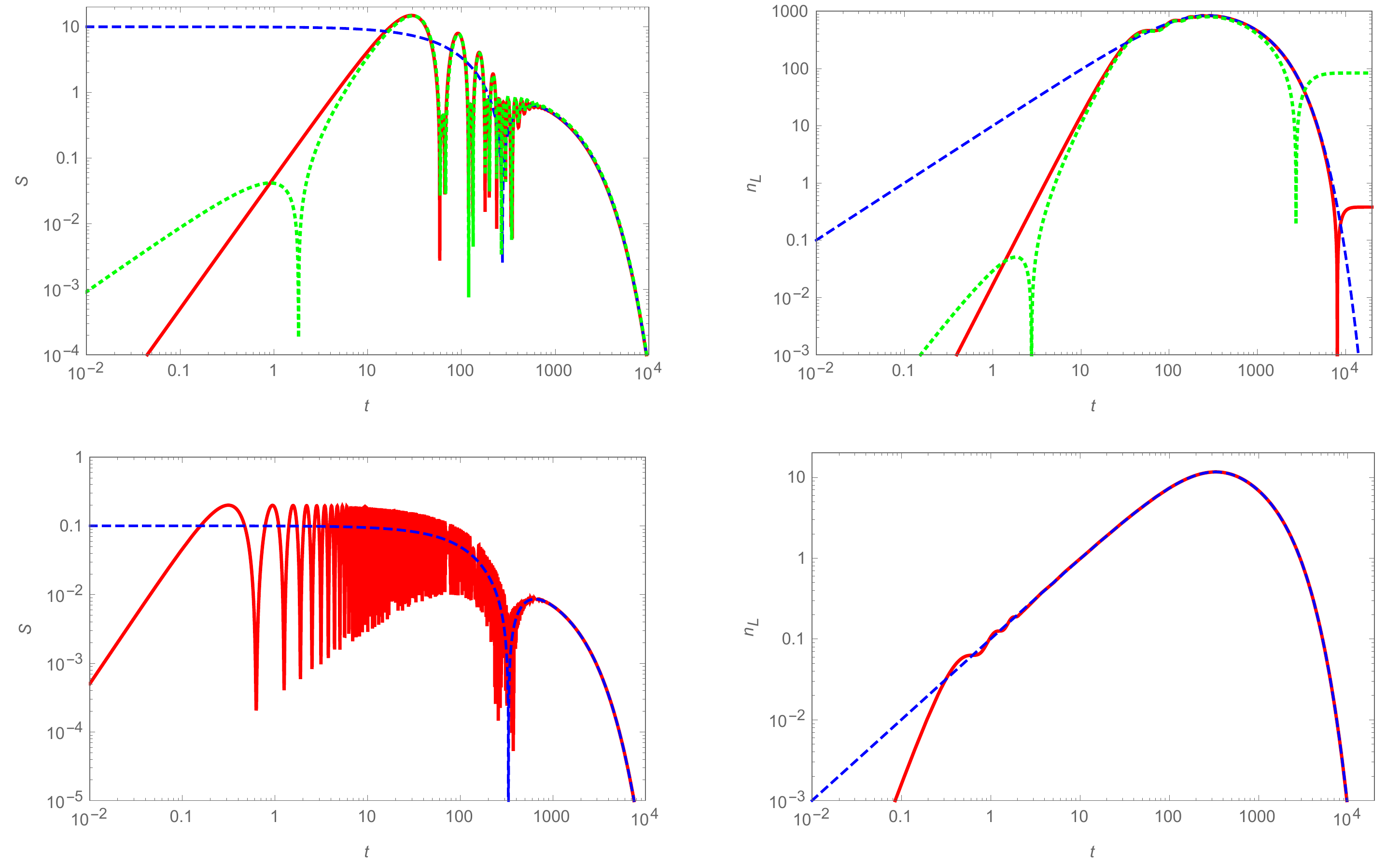,width=0.95\textwidth,angle=0}}} 
\caption[]{Source term (left plots), and lepton asymmetry (right plots), as a function of time normalized to $\gamma/\mmed$, with $\Gamma_1/\mmed=1/100$, $\Gamma_2/\mmed=1/120$, and $\Delta M/\mmed=0.1\, (10)$ in the top (bottom) plots. The lepton asymmetry has been obtained integrating only the source term (washouts are not considered). The solid red line corresponds to the source $S(T)$ from eq.~\ref{eq:source}, the dashed blue line to the source of the classical BE in the hierarchical limit (eq.~\ref{eq:sourcecl}), and the green dotted one  to the source $S(t)$ without including the interference terms (in the bottom plots the green curves are not shown because they are almost identical to the red ones). The scale on the vertical axis is not relevant and we have taken, for the purpose of illustration, $n^{\rm eq}(t)=e^{-\mmed \, t/(1000 \, \gamma)}$ (after a change of variables in the integration over time, the factor $\mmed/\gamma$ becomes part of the normalization chosen for the lepton asymmetry).} 
\label{fig:1}
\end{figure}
To discuss the limit of large mass splittings and the effect of the interference terms, we plot in  figure~\ref{fig:1} some variants of the source terms and the corresponding lepton asymmetries as a function of time for two different mass splittings, $\Delta M/\mmed=0.1$ (top plots) and a much larger value, $\Delta M/\mmed=10$, in the bottom plots. The decay widths have been chosen equal to $\Gamma_1/\mmed=1/100$ and $\Gamma_2/\mmed=1/120$. Finally, a larger time scale is chosen for the evolution of the number density $n^{\rm eq}$, namely we take $n^{\rm eq}(t)=e^{-\mmed \, t/(1000 \, \gamma)}$. In this way the time scales associated to oscillations, decays and equilibrium are well separated. The lepton asymmetry (right plots) has been obtained by integrating the source term over time, without considering any washouts. In this case unitarity requires that the final asymmetry be null. Indeed, this behavior can clearly be seen in the plots for the lepton asymmetry derived from $S^{\rm cl}(T)$ and also from $S(t)$ within the limit of our approximations. In this regard and according to the discussion at the end of section~\ref{sec:osc}, note that the final lepton asymmetry is much closer to zero if the full source term~\ref{eq:source} is considered, than if the interference terms are dropped. The plots also show that, as the mass splitting increases, the interference term becomes irrelevant and, moreover, the lepton asymmetry obtained from $S(t)$ tends to the one obtained from $S^{\rm cl}(T)$ at all times.  Note, however, that the probabilities obtained from the quantum field theory model of oscillations are valid up to first order in $(M_2^2-M_1^2)/(2 \pmed^2)$, so they cease to be valid for large mass splittings, for which actually oscillations are not expected to occur at all (and even within the range of validity of eqs.~\ref{eq:AL}, some decoherence factors we have neglected may become relevant).


\section{Conclusions and outlook}
\label{sec:con}
We have studied the sources of CP violation in a scalar toy model for baryogenesis with quasi-degenerate neutrinos. Our approach has been to use the renormalized propagator, diagonalized at first order in $\eta_{ij}$ (eq.~\ref{eq:eta}), in a quantum field theory model of neutrino oscillations. The probabilities for lepton number violating processes that we obtain (eqs.~\ref{eq:AL}) are valid up to first order in $(M_2^2-M_1^2)/(2 \pmed^2)$ and are compatible with unitarity up to sixth order in the Yukawa couplings (eq.~\ref{eq:unit3}). From these probabilities, which only involve the stable (anti)leptons in the initial and final states, we derived a source term for the evolution of the lepton asymmetry via a suitable time integral over the history of the system (eq.~\ref{eq:source}), without performing any subtraction of real intermediate states. This source term has contributions that can be identified with CP violation from mixing, oscillations and interference between both. In equilibrium the terms coming from CP violation in mixing and oscillations cancel, yielding a null source term as required by unitarity and CPT invariance.  Comparing with the standard classical approach, appropriate for non-oscillating neutrinos, we argued for a correspondence between the contribution from real intermediate state subtracted rates in the case of large mass splittings and the contribution from CP violation in oscillations in eq.~\ref{eq:source}. The interference terms between mixing and oscillations give  sub-dominant contributions of eighth order in the Yukawa couplings when integrated over all times.  However, at small times compared to the oscillation period, the interference terms are relevant and cancel the mixing contribution. Moreover, the effect of the interference terms becomes more important with increasing degeneracy, suggesting that they might have an important role in the highly degenerate limit, as found in~\cite{Kartavtsev15}. 

 We have performed this first study in a simple scalar toy model, within a static universe, and for a trivial momentum distribution of the particles. However, it is possible to extend the approach to an expanding universe and spin 1/2 neutrino fields with realistic momentum distributions, as well as to other type of scattering processes, in order to make a closer connection to ARS and resonant leptogenesis, including the intermediate mass regime. It can also be interesting to address the highly degenerate case, $M_2 - M_1 \sim \Gamma_{1,2}$, with the same point of view.

\section*{Acknowledgments}
We wish to thank Esteban Roulet for useful comments.

\bibliographystyle{JHEP}
\bibliography{referencias_leptogenesis3}

\end{document}